# Assessing contribution of treatment phases through tipping point analyses via counterfactual elicitation using rank preserving structural failure time models

Sudipta Bhattacharya[a]* and Jyotirmoy Dey[b]

[a]Statistics and Quantitative Sciences, Takeda, Cambridge, MA, USA;

[b]Haematology, Regeneron, NY,USA

*Corresponding Author : Sudipta Bhattacharya

e-mail Address: sudipta.bhattacharya@takeda.com

LinkedIn Sites of the Authors:

Sudipta Bhattacharya: https://www.linkedin.com/in/sudipta-bhattacharya-2498413b/

Jyotirmoy Dey: https://www.linkedin.com/in/jyotirmoy-dey-7a556014/

# Assessing contribution of treatment phases through tipping point analyses via counterfactual elicitation using rank preserving structural failure time models

Abstract: This article provides a novel approach to assess the importance of specific treatment phases within a treatment regimen through tipping point analyses (TPA) of a time-to-event endpoint using rank-preserving-structural-failure-time (RPSFT) modelling. In oncology clinical research, an experimental treatment is often added to the standard of care therapy in multiple treatment phases to improve patient outcomes. When the resulting new regimen provides a meaningful benefit over standard of care, gaining insights into the contribution of each treatment phase becomes important to properly guide clinical practice. New statistical approaches are needed since traditional methods are inadequate in answering such questions. RPSFT modelling is an approach for causal inference, typically used to adjust for treatment switching in randomized clinical trials with time-to-event endpoints. A tipping-point analysis is commonly used in situations where a statistically significant treatment effect is suspected to be an artifact of missing or unobserved data rather than a real treatment difference. The methodology proposed in this article is an amalgamation of these two ideas to investigate the contribution of a specific component of a regimen comprising multiple treatment phases. We provide different variants of the method and construct indices of contribution of a treatment phase to the overall benefit of a regimen that facilitates interpretation of results. The proposed approaches are illustrated with findings from a recently concluded, real-life phase 3 cancer clinical trial. We conclude with several considerations and recommendations for practical implementation of this new methodology.

.

Keywords: tipping point, time to event, survival analysis, Cox regression, counterfactual events

**Introduction**

Adding a new experimental drug to standard of care (SOC) therapy with the hope of improving efficacy is common practice in medical research. A randomized clinical trial (RCT) is usually needed to evaluate the efficacy and safety of the new regimen, comprising the new drug and the SOC, compared to SOC alone. When the experimental drug is utilized in multiple phases within the same regimen, it raises questions about the necessity of each phase, even if the overall regimen outperforms the SOC. In such cases, the effect of component phases can be confounded in ways that make their contribution to the overall efficacy of the regimen difficult to discern. There are several examples of RCTs with such component treatment phases where one experimental arm is compared to one SOC control arm. Stupp et al. (2005) published results of a study EORTC-22981/26981/NCIC-CE.3 in newly diagnosed glioblastoma multiforme patients who were administered a new drug temozolomide in combination with SOC radiation therapy (RT) followed by 6 cycles of maintenance with temozolomide. The study showed significant improvement in overall survival compared to SOC RT alone. Schmid et al. (2020) recently reported results of KEYNOTE-522, a study of pembrolizumab in *combination with neoadjuvant* SOC chemotherapy (CT) followed by *adjuvant* pembrolizumab compared to SOC in patients with early triple-negative breast cancer. The study showed significant improvement in recurrence-free survival (RFS) on the pembrolizumab arm, but the sponsors initially failed to secure approval of the United States Food and Drug Administration (US FDA) based on the RFS benefit until additional supportive overall survival (OS) data became available. Similarly, a recently reported phase 3 clinical trial in breast cancer patients, BROCADE3, by Diéras et al. (2020)

compared the efficacy of a new drug veliparib added to SOC CT during a (chemo-) combination phase followed by veliparib monotherapy at a higher dose in a maintenance phase when CT is discontinued. Since other drugs in the same class have demonstrated excellent efficacy only with maintenance therapy among those who responded to SOC-CT, a controversy has emerged regarding the contribution of the combination phase to the benefit the full regimen. BROCADE3 will be the illustrative example for our proposed method in this article.

Presenting our problem statement in general terms, when a new therapy A is added to standard of care therapy C to form a new treatment regimen for a given disease, the most reliable way to evaluate its efficacy is to conduct a randomized controlled trial (RCT) of A+C vs. C. However, when two new therapies A and B are added on to C, it is not adequate to show that A+B+C is more efficacious by simply comparing A+B+C vs C in a two-arm RCT, since it is generally not possible in this case to isolate the contribution of either A or B without making additional assumptions about the effect of each, or borrowing such information from external data. If there is scepticism about the contribution of A to the efficacy of A+B+C for example, it would be best assessed that effect in an RCT of A+B+C vs. B+C.

[Figure 1 to be inserted here.]

Figure 1B shows schematics of a full factorial design including treatment arms A+C and B+C that is ideal for isolating contribution of components while adding A and B to a standard of care C. In the specific context we have in mind, A and B represent treatment with the same experimental drug administered in two separate phases following different dosing amounts or schedules. As is usual in clinical research, these designs depict A as a treatment administered together with the SOC therapy C in the combination phase, and B is administered subsequently in the maintenance phase after discontinuation of C. Figure 1A depicts the general design schematics of A+B+C vs. C following the real-life trials EORTC-22981/26981/NCIC-CE.3, KEYNOTE-522 and BROCADE3 noted above.

In this article we propose a novel tipping point analysis (TPA) using rank preserving structural failure time (RPSFT) modelling that may be useful in assessing the influence of either component A or B on a time to event endpoint when they are administered in temporally separated treatment phases and as shown in the design schematics in Figure 1A. The method is illustrated by assessing the contribution of an experimental drug, veliparib, administered during the combination phase in the BROCADE3 example. In subsequent sections, we describe the mathematical framework for our proposed TPA methodology in detail (Section 2), provide published details of the study design and findings of the real-life example (Section 3), illustrate our proposed method by applying it to the example and exploring two variations estimating two different effects of interest (Section 4). We then conclude by discussing the pros and cons of implementing the method in Section 5, and provide practical recommendations and considerations for its use.

**Methodology**

Our proposed method adopts the following common framework for analysis of a time-to-event endpoint (TTE): Let $T$ denote the time from randomization to onset of the event of interest. In the context of clinical research, observations of such TTE variables are typically right-censored. If the censoring time is $R$, the observed outcome variables can be described as the right-censored TTE variable $S = \min(T, R)$, along with its censorship indicator $\Delta = I(T \leq R)$, where min(.) is the minimum function and I(.) is the indicator function. We will refer to $(S, \Delta)$ as the *TTE analysis doublet* in this article.

Next, consider a two-arm RCT of an experimental treatment E versus a control treatment C. Let $T_C$, $R_C$ and $(S_C, \Delta_C)$ respectively denote the uncensored TTE, censoring time and censored TTE analysis doublet for a subject receiving control therapy. Denote the same variables for a subject receiving the experimental therapy as $T_E$, $R_E$ and $(S_E, \Delta_E)$. We now assume that $T_C \sim F_C$ and $T_E \sim F_E$, with corresponding survival functions denoted by $S_C = 1 - F_C$ and $S_E = 1 - F_E$. For a well conducted study, $R_C$ and $R_E$ are generally governed by independent stochastic processes and assumed to be independent of $T_C$ and $T_E$ (either unconditionally or conditional on a set of baseline covariates), as well as the effect of the experimental treatment. When most of the censoring occurs due to administrative reasons (e.g., data cut-off for analysis), $R_C$ and $R_E$ are also be assumed to follow a common distribution $G$ independent of $S_C$ and $S_E$.

### *Traditional Approach using Cox Regression*

The Cox proportional hazards (PH) model has been the mainstay in analysing TTE data for as long as it has existed. It is the method of choice for drawing inference about the effect of specific factors, treatment or otherwise, that influence a TTE endpoint, Let us then consider the use of Cox regression first and outline why such an approach cannot provide satisfactory solutions for the problem at hand.

A Cox PH model fitted to all data from a trial of E (i.e., A+B+C) vs. C can be written as $S_E(t) = [S_C(t)]^\theta$. Let us also consider a Cox model with time-dependent covariates $I_A(.)$ and $I_B(.)$ that are indicators of time-periods prior to and following initiation of maintenance (with B or placebo) respectively. In its simplest form, this model can then be written as:

$$S_E(t) = [S_C(t)]^{\theta_1 I_A(t) + \theta_2 I_B(t)}.$$

To better understand what this model is estimating, suppose $S_E^A$ and $S_C^A$ denote survival functions for the experimental and control arms for the time interval between randomization and the onset of maintenance therapy or an event, whichever occurs earlier. We then have $S_E^A(u) = \left[S_C^A(u)\right]^{\theta_1}$, i.e., $\theta_1$ is the HR between the treatment arms during this interval. Similarly, we obtain $S_E^B(v) = \left[S_C^B(v)\right]^{\theta_2}$ where $S_E^B$ and $S_C^B$ are survival functions for the experimental and control arms for time from onset of maintenance therapy to an event among subjects who ultimately receive maintenance therapy (with possibly different populations for the two arms), and $\theta_2$ is the HR for that emerges between these two treatment arms during the maintenance phase, conditional on the different combination-phase treatment received prior to it. Thus, if one splits the dataset into two mutually exclusive subsets where the TTE observations for all subjects who received maintenance therapy is censored at the onset of that phase in the first, and only the TTE observations starting from the onset of maintenance is retained in the second (for subjects who received maintenance therapy), then simple Cox regression with treatment arm as covariate of these two subsets would yield the same estimates of $\theta_1$ and $\theta_2$, respectively, that the model with time-varying covariates shown above would.

We ask the reader now to note that, given the original study design (A+B+C vs. C), it is not possible to isolate effects of the two phases using the model shown above. It cannot account for any carryover effect of A emerging after the onset of maintenance without arbitrarily assigning a washout period for the effect of A (i.e., arbitrarily extending the positive support of $I_A$). It is confounded by the onset of B. Also, it does not account for a potentially delayed effect of B without arbitrarily shortening the positive support of $I_B$. Nor does it account for the potential differences in populations of subjects who transition to maintenance therapy on the two arms. As such, $\theta_2$ does not necessarily represent the pure effect of maintenance with B. And since comparative measurement of time-to-event during the combination phase is censored in the model when subjects initiate maintenance, $\theta_1$ cannot assess the full treatment effect of adding A during the combination phase.

*Tipping Point Analysis by Counterfactual Elicitation using RPSFT Modelling*

In causal inference of TTE endpoints in RCTs, RPSFT modelling is typically used to account for the presence of treatment switching. It provides a method for estimating and simulating counterfactual observations of the TTE under no switching ($T'$) by modifying the actual observed TTE under switching ($T$) through an adjustment factor $\lambda$. Let $X$ denote the time to onset of the ICE from randomization. If $Y = T - X$, then one can represent the counterfactual TTE observation under no switching as:

$$T' = X + \lambda Y$$

In modelling, when interest lies in estimating $T'$, one attempts to estimate the factor $\lambda$ and uses it to remove the influence of the ICE from $T$ (Robins and Tsiatis 1991; White et al. 1999; White et al. 1997). When the event of interest is an undesirable clinical outcome (such as progression of a disease), a positive effect of the ICE (treatment switching) would imply $\lambda < 1$ and $T > T'$, i.e., time to the event would be delayed (or prolonged) by switching to a more efficacious treatment. Conversely, $\lambda > 1$ and $T < T'$ would represent a negative effect of the ICE. As a convention, for subjects who do not experience the ICE, we will set $Y = 0$ and $X = T$ (the TTE). Within our proposed methodology, we will refer to this as the RPSFT structure and use it to generate counterfactual observations from hypothetical treatment arms using the observed trial data. We then pair this RPSFT modelling structure with a tipping point analysis to accomplish our inferential goals.

Tipping point analysis (Permutt 2016; Zhao et al. 2016) is an approach for assessing the impact of missing observations on statistically significant findings of an RCT when there is reason to suspect that those results are substantially influenced by the missingness. The approach works by imputing data for missing observations in a manner that is progressively conservative (i.e., biased against the experimental arm) until statistical significance is lost. The amount of conservatism is controlled by one or more parameters, and the parameter setting at which the imputation-based treatment difference crosses a specified threshold (e.g., become non-significant) is then referred to as the "tipping point".

The method proposed in this paper combines elements from these two techniques by first using the RPSFT structure in a natural way to account for the influence of component phases within a treatment regimen on study outcomes, and then performing a TPA using that structure to infer about the contribution of the specific treatment phase of interest. In describing our approach, we will hereafter denote by C, A+C, B+C and A+B+C the corresponding treatment/regimen as well as a study arm comprising that treatment/regimen.

Let us once again consider the RCT design depicted in Figure 1A where each subject's treatment consists of a combination phase followed by a maintenance phase. On the experimental arm, treatment A is given in combination with C during the combination phase, followed by treatment B in the maintenance phase. On the control arm, patients receive only C during the combination phase followed by placebo maintenance. Our inferential interest then lies in two distinct treatment effects:

Effect 1: The contribution of A to the effect of the full regimen (A+B+C); and

Effect 2: The sole effect of adding A to C (i.e., without the use of B in the regimen).

Effect 1 is best estimated from an RCT comparing A+B+C to B+C and Effect 2 is best estimated from an RCT comparing A+C to C (see Figure 1B). Since our actual design (Figure 1A) does not include either an A+C or a B+C arm, we will simulate counterfactual outcomes from such arms as described below.

*Model set-up and assessment method for Effect 1*

For estimation of Effect 1 we propose to simulate counterfactual observations from a hypothetical treatment arm B+C by modifying observations on the original control arm (C only) using the RPSFT structure, while leaving observations on the experimental arm A+B+C unchanged. The RPSFT model here postulates that the adjusted time to event for subjects on the hypothetical control treatment B+C may be obtained as:

$$T'_C = X_C + \lambda_C(T_C - X_C) = X_C + \lambda_C Y_C.$$

Since censorship is assumed to follow an independent distribution *G*, censoring times (observed or unobserved) would remain unaffected in the counterfactual. For assessing Effect 1, note that the adjustment factor $\lambda_C$ only applies to control-arm subjects who actually received maintenance therapy with placebo, i.e., for those with observed $Y_C > 0$. Observations on the experimental arm are left unaltered. When the experimental regimen A+B+C shows a benefit that is (hypothesized to be) primarily due to B, one assumes $\lambda_C \geq 1$.

Let $\theta(\lambda_C)$ denote the estimated hazard ratio (HR) between the experimental arm and the hypothetical control arm in the simulated sample; $p(\lambda_C)$ be the one-sided p-value for the test of significance of $\theta(\lambda_C)$; and $\theta_2(\lambda_C)$ be the estimated hazard ratio between the experimental and the hypothetical control arm for the maintenance phase (coefficient of $I_B$ in a corresponding Cox regression). Suppose $\theta^*$ and $\theta_2^*$ are the originally observed hazard ratios between the treatment arms (overall and during the maintenance period) of the study, and $p^*$ be the original overall p-value. In this formulation, $\lambda_C$ alone controls the counterfactual effect of the hypothetical control arm B+C and we can progressively increase it until a certain "tipping point" is reached. For our approach, we naturally require that $\theta(1) = \theta^* < 1$ and $p(1) = p^*$. We propose calculating tipping points using the following criteria:

(a) A value of $\lambda_C \geq 1$ that leads to loss of statistical significance of the treatment difference between the two arms. This follows the standard "tipping point" approach, If the corresponding tipping point is denoted $\lambda_{Ca}^*$, it is then defined as $\lambda_{Ca}^* = inf\{\lambda_C : p(\lambda_C) \geq 0.025\}$. It is easy to see that $\theta(\lambda_{Ca}^*)$ is approximately equal to the minimum detectable difference (MDD) of the study. The higher the estimate of $\lambda_{Ca}^*$, the more likely it is that any statistical significance between arms of the actual study design is not solely an effect of maintenance with B.

(b) A value of $\lambda_C \geq 1$ that "neutralizes" the treatment difference observed during the maintenance phase of the study. We define the tipping point as $\lambda_{Cb}^* = inf\{\lambda_C : \theta_2(\lambda_C) \geq 1\}$. Since $\lambda_{Cb}^*$ removes the treatment difference emerging after initiation of B that may include some carryover effect of A, $\theta(\lambda_{Cb}^*)$ provides an assessment of the minimum effect that can solely be ascribed to the addition of treatment A to the control C.

(c) A value of $\lambda_C \geq 1$ that "neutralizes" all treatment differences between the two arms. We define this tipping point as $\lambda_{Cc}^* = inf\{\lambda_C : \theta(\lambda_C) \geq 1\}$. The higher the value of $\lambda_{Cc}^*$, especially compared to $\lambda_{Cb}^*$, the more likely it is that treatment difference between arms of the actual study is not solely an effect of B.

All three of these tipping points provide an assessment of the contribution of the combination phase to the full regimen in distinct ways.

We now explain how counterfactual observations from the hypothetical experimental arm B+C shown in Figure 1B may be simulated based on the observed data for the control arm C in the original trial (Figure 1A). The following proposition formalizes that censored observations on the control arm C will remain unchanged by our proposed approach for estimating effect 1.

**Proposition 1a:** For the *i*-th subject on the control arm who receives placebo maintenance, suppose the observed TTE analysis doublet is $(s_{Ci}, \delta_{Ci})$ and the subject's time to initiating placebo maintenance, time to censoring and time from initiating maintenance therapy to an event (unobserved when $\delta_{Ci} = 0$) are $x_{Ci}, r_{Ci}$ and $y_{Ci}$, respectively. Then, for any given value of $\lambda_C \geq 1$, the subject's counterfactual TTE analysis doublet $(s'_{Ci}, \delta'_{Ci})$, can be obtained as:

$$(s'_{Ci}, \delta'_{Ci}) = (t'_{Ci}, 1) \quad \text{if } \delta_{Ci} = 1 \text{ and } t'_{Ci} = x_{Ci} + \lambda_C y_{Ci} \leq r_{Ci};$$

$$= (r_{Ci}, 0) \quad \text{if } \delta_{Ci} = 0; \text{ or if } \delta_{Ci} = 1 \text{ and } t'_{Ci} > r_{Ci}.$$

**Proof:** Given that $\lambda_C \geq 1$ we find $t'_{Ci} = x_{Ci} + \lambda_C y_{Ci} \geq x_{Ci} + y_{Ci} = t_{Ci}$.

When $\delta_{Ci} = 0$, we get $t'_{Ci} \geq t_{Ci} > r_{Ci}$, and one derives:

$$\delta'_{Ci} = I(t'_{Ci} \leq r_{Ci}) = I(t_{Ci} \leq r_{Ci}) = \delta_{Ci} = 0;$$

$$s'_{Ci} = \min(t'_{Ci}, r_{Ci}) = \min(t_{Ci}, r_{Ci}) = s_{Ci} = r_{Ci}.$$

When $\delta_{Ci} = 1$, $x_{Ci}$, $y_{Ci}$ and the actual time to event $t_{Ci} = x_{Ci} + y_{Ci}$ would have been observed. Since $\lambda_C \geq 1$ and $t'_{Ci} \geq t_{Ci}$, the hypothetical observation is observed if $t'_{Ci} \leq r_{Ci}$ else the observation is censored.

∎

Figure 2 provides a schematic elicitation of the counterfactual events generated in effect 1.

**[Figure 2]**

When an original observation is not censored, we propose to simulate its unobserved censoring time $r_{Ci}$ conditional on the fact that $t_{Ci} \leq r_{Ci}$. This conditioning fulfils the requirement that when B does not have an effect, i.e., when $\lambda_C = 1$, the original observations on C will remain unaltered and $\theta(\lambda_C) = \theta(1) = \theta^*$.

There are essentially two ways of simulating $r_{Ci}$ from this conditional distribution:

1. For a well-conducted study with low loss-to-follow-up rates, where a majority of censoring occurs due to data cutoff for analysis, it is reasonable to assume that the counterfactual disease progression or death of an originally uncensored subject would still have been observed if the counterfactual event were to occur by the data cutoff date. In other words, to construct the hypothetical survival doublet for analysis, $r_{Ci}$ may simply be imputed using the time from randomization to the data cutoff date when it is unobserved.

2. If censoring due to reasons other than administrative ones (e.g., data cutoff) are non-negligible, then it is better to obtain an estimate of the common, independent censorship distribution $G$ by fitting an appropriate parametric or semiparametric survival model to the study data after reversing the censorship indicator. This reduces simulation of the censoring times to the simple task of sampling $r_{Ci}$ from the fitted model using rejection sampling conditional on $r_{Ci} \geq t_{Ci}$.

Once the unobserved censoring times $r_{Ci}$ have been simulated, for any given value of $\lambda_C$ the counterfactual TTE analysis doublet can be generated using Proposition 1a.

*Model set-up and assessment method for Effect 2*

When interest lies in Effect 2, we propose to simulate counterfactual observations from a hypothetical treatment arm A+C through modification of the real observations on the experimental arm (A+B+C), leaving observations on the control arm unchanged. In this case, the RPSFT model is set up as follows:

$$T'_E = X_E + \lambda_E Y_E.$$

Since maintenance with treatment B is believed to be efficacious, in this case one assumes $0 < \lambda_E \leq 1$. In other words, the TTE endpoint is shortened if B is replaced by placebo on the experimental arm in the maintenance phase. We can now progressively decrease the factor $\lambda_E$ until a tipping point is crossed. Like before, we define our choices for the tipping point in this case as $\lambda_{Ea}^* = sup\{\lambda_E : p(\lambda_E) \geq 0.025\}$, $\lambda_{Eb}^* = sup\{\lambda_E : \theta_2(\lambda_E) \geq 1\}$ and $\lambda_{Ec}^* = sup\{\lambda_E : \theta(\lambda_E) \geq 1\}$. Also, all three of these tipping points provide an assessment of the contribution of the combination phase to the full regimen in distinct ways as before.

For assessing Effect 2, note that the adjustment factor $\lambda_E \leq 1$ only applies to subjects on the experimental treatment arm, who originally received maintenance therapy with B (those with observed $Y_E > 0$). Observations on the control arm are left unaltered. Similar to Proposition 1a, the following proposition formalizes how the TTE analysis doublet can be constructed in this case.

**Proposition 1b:** For the $i$-th subject on the experimental arm who receives maintenance with B, let the observed TTE analysis doublet be $(s_{Ei}, \delta_{Ei})$ with time to initiating maintenance $x_{Ei}$, time to censoring $r_{Ei}$ and time from initiating maintenance to an event $y_{Ei}$ (possibly unobserved). Let $t'_{Ei} = x_{Ei} + \lambda_E y_{Ei}$. Then, given any $0 < \lambda_E \leq 1$, the counterfactual TTE analysis doublet can be obtained as:

$$(s'_{Ei}, \delta'_{Ei}) = (t'_{Ei}, 1) \quad \text{if } \delta_{Ei} = 1; \text{ or if } \delta_{Ei} = 0 \text{ and } \lambda_E y_{Ei} \leq r_{Ei} - x_{Ei};$$
$$= (r_{Ei}, 0) \quad \text{if } \delta_{Ei} = 0 \text{ and } \lambda_E y_{Ei} > r_{Ei} - x_{Ei}.$$

**Proof:** Given $\lambda_E \leq 1$, observe that $t'_{Ei} = x_{Ei} + \lambda_E y_{Ei} \leq x_{Ei} + y_{Ei} = t_{Ei}$ and, since $t'_{Ei} \leq t_{Ei} \leq r_{Ei}$ for $\delta_{Ei} = 1$, we have $(s'_{Ei}, \delta'_{Ei}) = (t'_{Ei}, 1)$. On the other hand, when $\delta_{Ei} = 0$, $r_{Ei}$ is observed but $t_{Ei}$ is censored (and one must impute $y_{Ei}$). If $\lambda_E y_{Ei} \leq r_{Ei} - x_{Ei}$, then $t'_{Ei} \leq r_{Ei}$ and the counterfactual TTE would also be observed, i.e., $\delta'_{Ei} = 1$. Otherwise, if $\lambda_E y_{Ei} > r_{Ei} - x_{Ei}$ then $t'_{Ei} > r_{Ei}$ and it remains censored, i.e., $\delta'_{Ei} = 0$, with $s'_{Ei} = r_{Ei}$.

∎

Figure 3 provides a schematic elicitation of the counterfactual events generated in effect 2.

[Figure 3]

Unobserved event times $T_E$ for subjects who receive maintenance with B and are censored need to be simulated conditional on $T_E > R_E$ to obtain their counterfactual observation on a hypothetical A+C arm. We can try to use an estimate of $F_E$ conditional on $T_E > R_E$ to impute the time to event if it is mathematically tractable. For example, if $F_E$ is an exponential distribution, then conditional on $T_E > R_E$ the distribution of $T_E$ remains the same due to its memoryless property. Our preferred approach, however, is to use a fitted survival model to time on maintenance with B and impute $T_E'$ conditional on $T_E > R_E$ using a rejection sampling scheme.

As the title of this section suggests, we refer to this new methodology as *Tipping Point Analysis by Counterfactual Elicitation (TPACE)*. Interpretation of a tipping point depends on its extremeness. The more extreme the tipping point is in terms of bias against the experimental therapy, the more unlikely it is that any effect ascribed to that therapy could potentially be a result of unobserved outcomes. This assessment is usually qualitative since there is typically no clear benchmark to determine how extreme a tipping point truly is. In the next section however, we describe how an informal interpretation of estimated Effects 1 and 2 may be obtained by calculating a contribution index involving the estimated tipping points obtained via TPACE.

*Interpretating TPACE-based estimates and assessing contribution of a treatment phase*

Given A+B+C is observed to be more effective than C, we have assumed that there exists a constant scaling $\lambda_C$ (>1) of the control arm C that neutralizes the treatment difference emerging during the maintenance period of the original study following the RPSFT structure in the context of assessing Effect 1 using TPACE. Using notation introduced in Section 2.1, we can denote this as $S_E^B(v) = \left[S_C^B(v)\right]^{\theta_2} \cong S_C^B\left(\frac{v}{\lambda_C}\right)$. When $S_E^B$ follows an exponential distribution, a common assumption in the modeling of TTE data in RCTs, we would surmise that $\lambda_C = \frac{1}{\theta_2}$ and $S_E^B(v) = S_C^B(v\theta_2)$.

If the observed TTE duration following initiation of maintenance therapy on the original control arm C is prolonged (and we do not concern ourselves with potential additional censoring when TTE is prolonged) using the scaling factor $\frac{1}{\theta_2}$ and $\theta'$ denotes the resultant HR, we can derive a relationship between $S_E$ and $S_C$ as:

$$S_E(t) \cong \left[S_C\left(x_C + \frac{t - x_C}{\theta_2}\right)\right]^{\theta'}.$$

Considering the time dilation shown above to be one way of neutralizing the effect of the maintenance phase $\theta_2$ by reducing it to nearly 1, one may view $\theta'$ as providing an assessment of the residual effect that can only be ascribed to the addition of A during the combination phase. As one might expect, this results in a value of $\theta'$ approximately equal to $\theta_1$. To realize this through counterfactuals one simply scales both event and censoring times during the maintenance phase by the same factor representing a treatment effect and rerunning the Cox model. However, it distorts the censoring distribution such that it is no longer independent of the TTE distribution and treatment effect, and therefore violates fundamental assumptions of the Cox model. In practice, when censoring occurs primarily due to length of follow-up, the above scaling is tantamount to selectively increasing the follow-up period for subjects initiating maintenance and in a manner that is dependent on the treatment effect. Hence, findings thereof are biased.

*Assessment index for Effect 1*

In TPACE we use proper counterfactual observations to avoid the distortion of the censorship distribution mentioned above and neutralize the effect of the component phases within the Cox model using the RPSFT structure. Thus, for Effect 1 we estimate the scaling factor $\lambda^*_{Cb} > 1$ that yields a counterfactual $\theta_2(\lambda^*_{Cb}) \cong 1$ and the scaling factor $\lambda^*_{Cc}$ such that we have counterfactual $\theta(\lambda^*_{Cc}) \cong 1$ using TPACE, while keeping the censoring distribution nearly unchanged (and independent). If now $T^E_C$ and $T^B_C$ respectively denote the estimates of the prolonged hypothetical TTE on the control arm that neutralize the overall difference between arms of the actual study (mimicking arm A+B+C) and the difference between treatment arms that emerged after onset of maintenance (mimicking arm B+C), we may write them as:

$$T^E_C = X_C + \lambda^*_{Cc} Y_C \quad \text{and} \quad T^B_C = X_C + \lambda^*_{Cb} Y_C.$$

We can then split the total estimated delay in TTE on the experimental arm (E = A+B+C) gained over C as the sum of the estimated delay gained after onset of maintenance $T^B_C - T_C$ and the delay gained prior to that (credited to treatment A), denoted $T^A_C$, i.e.,

$$T^E_C - T_C = (T^E_C - T^B_C) + (T^B_C - T_C) = T^A_C + (T^B_C - T_C).$$

We thus obtain:

$$\text{Estimated time gained with E over C} = T^E_C - T_C = (\lambda^*_{Cc} - 1) Y_C,$$

$$\text{Time gained following onset of B} = T^B_C - T_C = (\lambda^*_{Cb} - 1) Y_C, \text{ and}$$

$$\text{Time gained due to A} = T^A_C = (T^E_C - T^B_C) = (\lambda^*_{Cc} - \lambda^*_{Cb}) Y_C.$$

We may now use the above expressions to derive the contribution index of A to the overall effect of E (i.e., A+B+C vs. B+C) as:

$$C^A_C = \frac{(T^E_C - T^B_C)}{(T^E_C - T_C)} = \frac{(\lambda^*_{Cc} - \lambda^*_{Cb})}{(\lambda^*_{Cc} - 1)}.$$

And the contribution index of B to overall effect of E is then the complement of $C_C^A$ derived as:

$$C_C^B = \frac{(T_C^B - T_C)}{(T_C^E - T_C)} = \frac{(\lambda_{Cb}^* - 1)}{(\lambda_{Cc}^* - 1)} = 1 - C_C^A.$$

Since the basic Cox model used in effect-neutralization does not account for effects of A carried over to the maintenance phase, $C_C^A$ should be viewed as an index of minimum contribution of A and $C_C^B$ as an index of maximum contribution of B to the total time gained by E over C.

*Assessment Index for Effect 2*

When using TPACE for Effect 2, we estimate the scaling factor $\lambda_{Eb}^* < 1$ such that counterfactual $\theta_2(\lambda_{Eb}^*) \cong 1$ and the scaling factor $\lambda_{Ec}^* < 1$ such that counterfactual $\theta(\lambda_{Ec}^*) \cong 1$, maintaining independence of the censoring distribution. If now $T_E^E$ and $T_E^B$ respectively denote the estimates of the shortened hypothetical TTE on the experimental arm that eliminates the overall difference between arms E and C (thus mimicking arm C) and the difference between treatment arms that emerged after onset of maintenance (thus mimicking arm A+C), we may write them as:

$$T_E^E = X_E + \lambda_{Ec}^* Y_E \quad \text{and} \quad T_E^B = X_E + \lambda_{Eb}^* Y_E.$$

The difference between the two $T_E^B - T_E$ then estimates the gain due to adding A alone to C (i.e., Effect 2) on the time scale. Like before, we can then obtain:

Estimated time gain of E over C eliminated = $T_E^E - T_C = (\lambda_{Ec}^* - 1) Y_E$, and

Estimated time gain following onset of B eliminated = $T_E^B - T_C = (\lambda_{Eb}^* - 1) Y_E$.

Therefore, estimated time gain only due to A = $T_E^A = (T_E^E - T_E^B) = (\lambda_{Ec}^* - \lambda_{Eb}^*) Y_E$.

We now derive the minimum individual efficacy index of A as:

$$E_E^A = \frac{(T_E^E - T_C^B)}{(T_E^E - T_C)} = \frac{(\lambda_{Ec}^* - \lambda_{Eb}^*)}{(\lambda_{Ec}^* - 1)}.$$

And similarly derive by $E_E^B = 1 - E_E^A$ the estimated maximum index of contribution of B to the full benefit of E when B is added to A+C.

**Application of TPACE**

The methodology proposed in this article is now applied to a phase 3 randomized, double-blinded, placebo-controlled study of veliparib, an inhibitor of the enzyme poly ADP ribose polymerase (PARP), in subjects with BRCA-mutated HER2-negative breast cancer, BROCADE3 (Diéras et al. 2020) In this RCT, subjects were randomized 2:1 to receive either veliparib or placebo added on to SOC chemotherapy (CT) with carboplatin and paclitaxel (Figure 4). While treatment with all three drugs (carboplatin, paclitaxel and veliparib/placebo) could continue until disease progression or unacceptable toxicity, study subjects had the option to discontinue any of these drugs at any time. When both SOC CT drugs were discontinued without disease progression or death, subjects also had the option to receive maintenance therapy with blinded study drug, veliparib at a higher dose or placebo.

[Figure 4]

The study showed a statistically significant improvement in progression-free survival (PFS) between the two treatment arms (Table 1, Figure 5). Subjects across both arms received 7.5 months of SOC CT on average (median 6.3 months) with 37% receiving veliparib monotherapy maintenance at a higher dose. A total of 349 subjects had experienced at least one PFS event by the time of the primary analysis and hence, the minimum detectable difference (MDD) at the 2-sided 5% level of significance can be easily calculated to be HR approximately equal to 0.8.

[Figure 5]

[Table 1]

It is important to assess the contribution of the combination phase to the overall treatment benefit of the veliparib regimen in BROCADE3 for several reasons. First, as noted in Diéras et al.(2020), the average duration of the combination phase was approximately 10 months and the survival distributions (KM curves) of the two treatment arms did not begin to separate meaningfully until approximately 12 months from randomization. However, subjects could receive combination therapy as long as it remained tolerable or until a PFS event occurred and substantial proportions of subjects continued their combination phase well past 12 months suggesting that the benefit of the experimental regimen may not be driven by maintenance alone. At the time of the primary analysis, as noted in Diéras et al.(2020), 38% of the 509 subjects included in the intent-to-treat (ITT) analysis had transitioned to veliparib/placebo maintenance,. Second, while studies of other PARP-inhibitors as maintenance therapy alone have demonstrated substantial efficacy leading to scepticism about the need for veliparib in the combination phase, such cross-study comparisons are inferentially fraught since studies of other PARPi were restricted only to subjects who had already received and responded to SOC CT prior to enrolment. Finally, while one cannot deem combination phase veliparib unnecessary for efficacy simply based on the above considerations, SOC CT is not without toxicity and adding veliparib to the combination phase further increases the treatment burden for study subjects.

To assess the contribution of combination-phase veliparib to the overall treatment effect, (Diéras et al. 2020). published results of a Cox regression analysis of the PFS endpoint including a time-varying covariate for initiation of veliparib/placebo maintenance, treatment assigned at randomization and their interaction (Table 2). The fitted model produced HR estimates for the combination phase effect $\theta_1$, 0.811 (95% CI: 0.622, 1.056), and the maintenance phase effect $\theta_2$, 0.493 (95% CI: 0.334, 0.728) favouring the veliparib arm.

[Table 2]

As described earlier, we view our motivating example of BROCADE3 as a study of A+B+C vs. C where C represents the SOC CT with paclitaxel and carboplatin, A is veliparib used as part of a combination phase with CT and B is veliparib monotherapy given at a higher dose during a maintenance phase after CT is discontinued (Figure 1).

*Effect 1: Indices of Contribution to PFS*

TPACE was performed to assess the contribution index of A in our example. Since the experimental regimen is more efficacious, a lower percentage of events was observed on the experimental arm compared to the control arm during the maintenance phase (Figure 3). Among all censored observations across both arms, PFS times for most subjects (64.5%) were censored at data cut-off. We therefore simulated the counterfactual observations to assess Effect 1 by simply imputing the time from randomization to data cut-off as the unobserved censoring time for subjects who progressed or died after transitioning to maintenance monotherapy. We then calculated the tipping points for PFS following criteria (a) (Figure 6A). (b) (Figure 6C) and (c) of our methodology (Table 3). The estimated loss of significance tipping point $\lambda^*_{Ca} = 1.79$ yields a "residual" counterfactual hazard ratio $\theta(1.79) = 0.802$. While the similarity of this HR to the MDD (0.805) is expected, its proximity to the Cox model's estimate of $\theta_1 = 0.811$ in our example however, is simply coincidental. The tipping point $\lambda^*_{Cb}$ is estimated to be 3.48 where the counterfactual treatment differences reduce to $\theta(3.48) = 0.908$ overall and $\theta_2(3.48) = 1$ in the period following initiation of maintenance. An estimated $\lambda^*_{Cc}=5.15$ fully neutralizes the full, overall difference between treatment arms with counterfactual $\theta(5.15) = 1$.

[Figure 6]

[Table 3]

Expressed in words, PFS following initiation of maintenance on the control arm needs to be extended by an additional 248% to get the hazard during this phase to match (on average) the hazard of the maintenance phase on the experiment arm. Since it needs to be extended by an unrealistic 415% to get the PFS for the control arm to match (on average) the experimental arm overall, the contribution index of A to the overall effect of E is calculated to be:

$$C_C^A = \frac{(5.15 - 3.48)}{(5.15 - 1)} = \frac{1.67}{4.15} = 0.402.$$

Thus, in words this suggests that, at least an estimated 40% of the total delay in disease progression or death observed on the experimental arm compared to control may be ascribed to the combination phase veliparib, and no more than 60% of the PFS prolongation after initiation of maintenance can be ascribed to maintenance veliparib alone.

### *Effect 2: Index for the Individual Efficacy of A*

As described in Section 2, TPACE for assessing Effect 2 was performed with the shrinkage factor $\lambda_E$ ($0 < \lambda_E < 1$) applied only to those who transitioned to maintenance on the experimental arm. Counterfactual PFS times for subjects who were randomized to the experimental arm and went on to receive maintenance veliparib needed to be simulated conditional on observed data in this case. We assumed and fit an exponential distribution for the subjects' time from initiation of maintenance to progression or death. For subjects censored during the maintenance phase, the unobserved time from censoring to potential progression or death would also follow the same exponential distribution by virtue of its lack-of-memory property. For these subjects, we therefore simulated the unobserved time from censorship to event using the fitted exponential model and added that time to the observed $s_{Ei}$ to obtain $t_{Ei}$. With the time to event ($t_{Ei}$ values) thus obtained for all subjects on the experimental arm, we shrink them by the factor $\lambda_E$ following the rules described in Section 2. For subjects who were originally censored during monotherapy if the shrunk, imputed time to event is smaller than the observed censoring time, the counterfactual observation for that subject will be an imputed event with PFS time set equal to the imputed time to event. Once again, starting from $\lambda_E = 1$ (where results are same as that of the primary analysis), grid-searches by decreasing the value of $\lambda_E$ were performed to find tipping points corresponding to thresholds (a), (b) and (c) for PFS (Figure 6B, 6D and Table 3).

Loss of statistical significance occurred at $\lambda^*_{Ea} = 0.82$ with counterfactual hazard ratio $\theta(0.82) = 0.811$. Once again this is simply approximating the MDD. More importantly, we estimated $\lambda^*_{Eb}$ to be 0.63, yielding the counterfactual $\theta(0.63) = 0.913$ and estimated $\lambda^*_{Ec}$ to be 0.48 yielding the counterfactual $\theta(0.48) = 1.001$. Finally, estimated the index of minimum individual efficacy of A as:

$$E^A_E = \frac{(0.48 - 0.63)}{(0.48 - 1)} = \frac{0.15}{0.52} = 0.288.$$

This suggests that if the option of maintenance veliparib was not part of the experimental arm, adding veliparib only to SOC CT in the combination phase would have achieved at least 29% of the total additional prolongation of PFS achieved by the experimental arm overall compared to SOC CT.

**Discussion**

Our proposed TPACE approach is an attempt to gain insights into the utility of one or more component phases within an efficacious experimental regimen comprising multiple phases. The study design shown in Figure 1A is suboptimal for isolating the effect of either A or B since the effect of B will always remain confounded with potential carryover effects of A. Ideally, one would need an RCT that includes study arms A+C or B+C to isolate the effect of A or B. These arms are sometimes not part of the actual study design due to practical reasons or study feasibility that limit the number of study subjects that can be recruited and studied in a reasonable amount of time. While TPACE cannot fully compensate for such design deficiencies, it provides a way to otherwise assess the utility of A within the implemented design by leveraging the temporal separation of the treatment phases.

In Section 2, two possible estimands, Effects 1 and 2, are described. Both assess the benefit that A provides to patients in different ways. The first corresponds to the contribution to A to the full A+B+C regimen obtained when it is added to B+C, estimated as the comparative effects of A+B+C vs B+C. The second corresponds to the individual efficacy of A when added to C, estimated as the comparative effect of A+C vs C. These two effects are of course, not the same as the former includes synergistic (or antagonistic) effects of A and B, which the latter does not. From the perspective of clinical practice and healthcare authorities, Effect 1 seems far more important to assess than Effect 2 since the study results only support the full regimen A+B+C as a future treatment for patients, and not A+C. Thus, while individually each drug (A or B) may only be modestly efficacious, together their benefit may be substantial for patients.

In view of the confounding of the effects of A and B in the original study design, TPACE or any other statistical method (e.g., Cox regression) are essentially only able to assess these effects conditional on the rest of the treatment regimen remaining unchanged. To interpret these as isolated, unconditional effects one would need to make strong simplifying assumptions such as:

   a. Any carryover effect of A is washed out prior to initiation of B and that the magnitude of such carryover effect is the same irrespective of the duration or actual dosage of A received,

b. Any effect of B begins from the day it is initiated and is the same irrespective of its duration or actual dosage received, and

c. There is little or no interaction between of the effects of A and B.

It is usually not possible to determine the validity of such assumptions in any conclusive manner based on the A+B+C study design. For example, the time to wash out of the carryover effect of A is usually intractable and cannot be reliably estimated from study data alone. It is possible to formulate alternate modelling structures assuming specific deviations from these assumptions and obtain effect estimates under such conditions. Under conventional approaches, such alternate modelling often comes with substantial added complexity and challenges in estimation and interpretation. Counterfactual elicitations in TPACE should prove comparatively simpler for implementing such alternate assumptions despite the limitation of not being able to completely eliminate treatment effect confounding.

Application of our method has been illustrated in this paper through assessment of the effect of A within the design schematic in Figure 1A. It is natural to ask if it is possible to assess the contribution of B to the full regimen in a similar way, should it be the focus of interest. The answer to this will depend on the design of the study. For example, this is not feasible in BROCADE3 since initiation of maintenance is dependent on the effect of combination therapy (on its efficacy or tolerability). There is no meaningful information in this context to generate counterfactual observations for a hypothetical A+C treatment arm from arm C observations for comparison with A+B+C. With some additional assumptions, a hypothetical B+C vs C comparison may however be formulated. The effect of B is however not in question in our illustrative example and hence we have considered this out of scope for this article.

In the context of treatment crossover, the Accelerated Failure Time (AFT) model is sometimes used in the same way as the RPSFT model for time to event endpoints to adjust for treatment switching (Latimer et al. 2014; 2017) It is easy to see that one can adapt the structure of the AFT model in the same way as we have used the RPSFT modelling structure to achieve similar inferential goals. Our recommendation then is, whichever model is chosen, one should employ it consistently for estimation of all the tipping points suggested to evaluate the proposed indices for inferential purposes.

Finally, we conclude with a few technical notes about our proposed methodology. First, we re-emphasize the importance of conditioning on the observed data (i.e., sampling from conditional distributions) when imputing time-to-event observations. This is critical to ensure a form of *effect anchoring* such that, when the observed dataset is left unchanged (i.e., the RPSFT model parameter $\lambda$ is set to 1), the effect estimates will match the actual findings of the original study. Second, for Effect 2, our approach is dependent on estimating and sampling data from the TTE distribution for subjects with censored observations. We advise paying close attention to the estimation and simulation steps during implementation of our method. When the proportion of observations censored is low but censoring is due to non-administrative reasons, it may be challenging to meaningfully impute potential censoring times as their distribution cannot be reliably estimated. Also, when the TTE distribution is influenced by nonproportional hazards (e.g., late separation), common parametric models may not provide a good fit to the observed data for the purposes of model-based imputation (like we have employed the exponential distribution in our illustrative application) and a bootstrapping approach may be simpler and preferable. There has been considerable interest in modelling TTE data under non-proportional hazards in recent years, and there may be alternate approaches to formulate TPACE with such modelling structures, which can be a topic of further research.

Acknowledgements: The authors are deeply thankful to Professor Gary Koch for his review of this research work and suggestions. The work has also greatly benefitted through discussions with and input from fellow research colleagues David Maag (Ph.D.) and Bruce Bach (M.D., Ph.D.).

**References**

Stupp, R., W.P. Mason, M.J. Van Den Bent, M. Weller, B. Fisher, M.J.B. Taphoorn, K. Belanger, A.A. Brandes, C. Marosi, U. Bogdahn, J. Curschmann, R.C. Janzer, S.K. Ludwin, T. Gorlia, A. Allgeier, D. Lacombe, J.G. Cairncross, E. Eisenhauer and R.O. Mirimanoff. 2005. Radiotherapy plus concomitant and adjuvant temozolomide for glioblastoma. *New England Journal of Medicine* 352, no 10: 987-96.

Schmid, P., J. Cortes, L. Pusztai, H. Mcarthur, S. Kümmel, J. Bergh, C. Denkert, Y.H. Park, R. Hui, N. Harbeck, M. Takahashi, T. Foukakis, P.A. Fasching, F. Cardoso, M. Untch, L. Jia, V. Karantza, J. Zhao, G. Aktan, R. Dent and J. O'shaughnessy. 2020. Pembrolizumab for early triple-negative breast cancer. 382, no 9: 810-21.

Diéras, V., H.S. Han, B. Kaufman, H. Wildiers, M. Friedlander, J.P. Ayoub, S.L. Puhalla, I. Bondarenko, M. Campone, E.H. Jakobsen, M. Jalving, C. Oprean, M. Palácová, Y.H. Park, Y. Shparyk, E. Yañez, N. Khandelwal, M.G. Kundu, M. Dudley, C.K. Ratajczak, D. Maag and B.K. Arun. 2020. Veliparib with carboplatin and paclitaxel in brca-mutated advanced breast cancer (brocade3): A randomised, double-blind, placebo-controlled, phase 3 trial. *Lancet Oncology* 21, no 10: 1269-82.

Robins, J.M. and A.A. Tsiatis. 1991. Correcting for non-compliance in randomized trials using rank preserving structural failure time models. *Commun. Stat. Theory Methods* 20, no 8: 2609-31.

White, I.R., A.G. Babiker, S. Walker and J.H. Darbyshire. 1999. Randomization-based methods for correcting for treatment changes: Examples from the concorde trial. *Statistics in Medicine* 18, no 19: 2617-34.

White, I.R., S. Walker, A.G. Babiker and J.H. Darbyshire. 1997. Impact of treatment changes on the interpretation of the concorde trial. *AIDS* 11, no 8: 999-1006.

Permutt, T. 2016. Sensitivity analysis for missing data in regulatory submissions. *Statistics in Medicine* 35, no 17: 2876-79.

Zhao, Y., B.R. Saville, H. Zhou and G.G. Koch. 2016. Sensitivity analysis for missing outcomes in time-to-event data with covariate adjustment. *Journal of Biopharmaceutical Statistics* 26, no 2: 269-79.

Latimer, N.R., K.R. Abrams, P.C. Lambert, M.J. Crowther, A.J. Wailoo, J.P. Morden, R.L. Akehurst and M.J. Campbell. 2014. Adjusting survival time estimates to account for treatment switching in randomized controlled trials--an economic evaluation context: Methods, limitations, and recommendations. *Medical Decision Making* 34, no 3: 387-402.

Latimer, N.R., K.R. Abrams, P.C. Lambert, M.J. Crowther, A.J. Wailoo, J.P. Morden, R.L. Akehurst and M.J. Campbell. 2017. Adjusting for treatment switching in randomised controlled trials - a simulation study and a simplified two-stage method. *Statistical Methods in Medical Research* 26, no 2: 724-51.

**Table 1:** Primary progression-free survival results of study BROCADE3

|  | **Experimental Arm (n=337)** | **Control Arm (n=172)** |
|---|---|---|
| PFS events | 217 | 132 |
| Censored observations | 120 | 40 |
| Median PFS (months) | 14.5 | 12.6 |
| Hazard ratio (95% CI)[#][*] | 0.71 (0.57, 0.88) | |
| Log-rank[*] p-value | 0.0016 | |

\# Hazard ratio estimated by fitting Cox proportional hazard (PH) model to the data.

\* Stratified by randomization stratification factors.

Generated with permission of Elsevier Pvt. Ltd., based on data provided in figure 2 and figure 3 of the article by Diéras et al. (2020)).

**Table 2**   Cox proportional hazards regression of progression-free survival with time-varying covariate indicating treatment phase.

|  | Combination phase | | Monotherapy phase | |
|---|---|---|---|---|
|  | Veliparib + C/P (N=337) | Placebo + C/P (N=172) | Veliparib + C/P (N=136) | Placebo + C/P (N=58) |
| Events/patients at risk at beginning of phase (%) | 149 / 337 (44%) | 88 / 172 (51%) | 68 / 131 (52%) | 44 / 57 (77%) |
| Hazard Ratio (95% CI) | 0.81 (0.62, 1.06) | | 0.49 (0.33, 0.73) | |

Treatment phase was included as a time-varying covariate which was set to 1 at and after initiation of monotherapy, and 0 prior to it.

(reprinted with permission of Elsevier Pvt. Ltd. from article by Diéras et al. (2020))

**Table 3:** Results of tipping point analysis evaluating effect of the combination phase.

| Effect / Estimated tipping parameter | Adjustment factor at tipping point | Average No. of Events | Hazard ratio at tipping point | P-value at tipping point[(a)] |
|---|---|---|---|---|
| Assessment for Effect 1 | | | | |
| $\lambda^*_{Ca}$ | 1.79 | 342 | 0.802[a] | 0.0503[#] |
| $\lambda^*_{Cb}$ | 3.48 | 335 | 0.908 | |
| $\lambda^*_{Cc}$ | 5.15 | 328 | 1.000 | |
| Assessment of Effect 2 | | | | |
| $\lambda^*_{Ea}$ | 0.82 | 379 | 0.811[a] | 0.0536 |
| $\lambda^*_{Eb}$ | 0.63 | 397 | 0.913 | |
| $\lambda^*_{Ec}$ | 0..48 | 406 | 1.001 | |

i. Note, the tipping value $\lambda^*_{Ca}$ and $\lambda^*_{Ea}$ are determined at the corresponding tipping parameters achieving minimum p > 0.05 (2-sided).

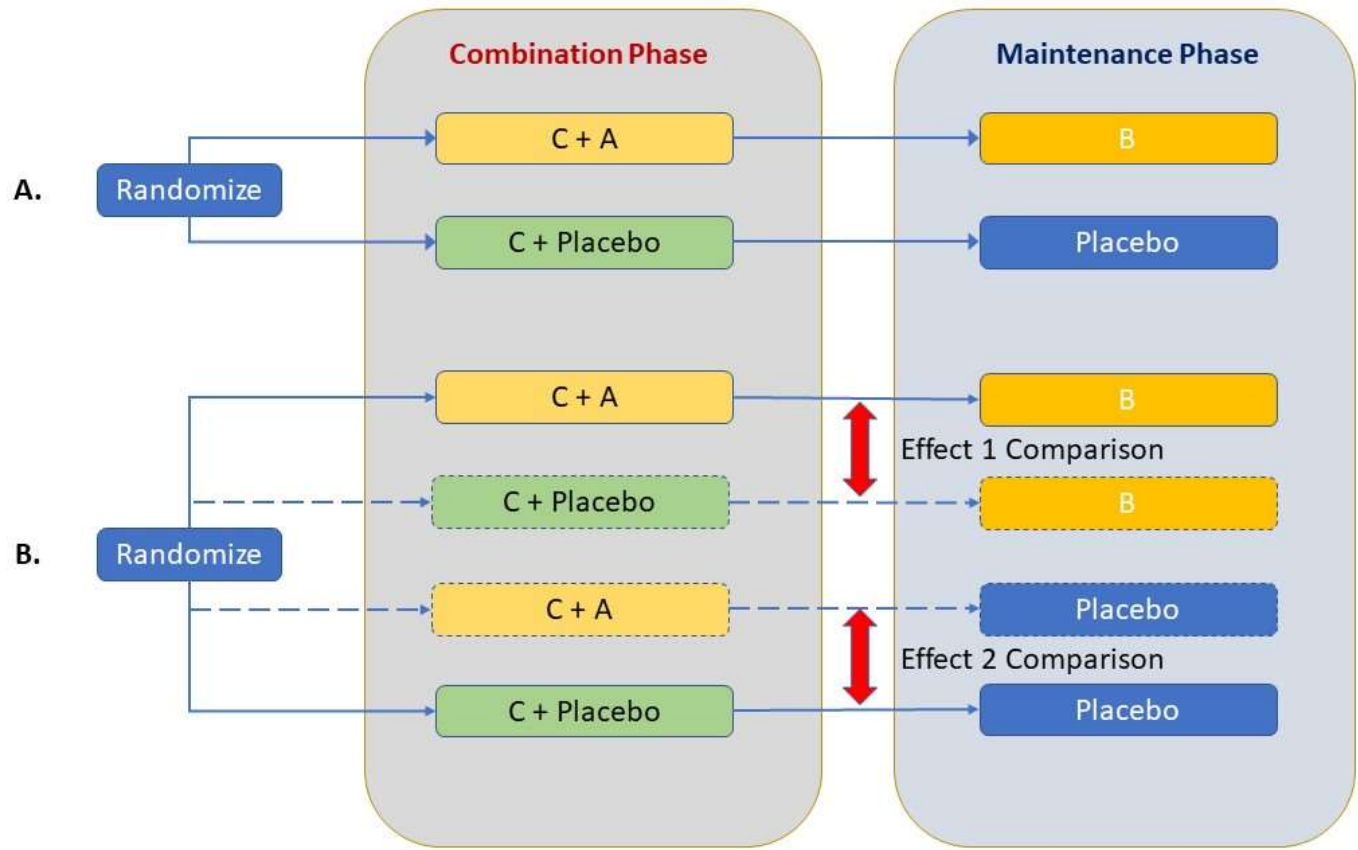

**Figure 1.** Design schematics of (**A**) a generic study with experimental treatment in both combination phase (treatment A) and maintenance phase (treatment B) to standard of care (treatment C) as in studies EORTC-22981/26981/NCIC-CE.3, KEYNOTE-522 or BROCADE3; and (**B**) a full-factorial design involving treatment A and/or treatment B added to treatment C.

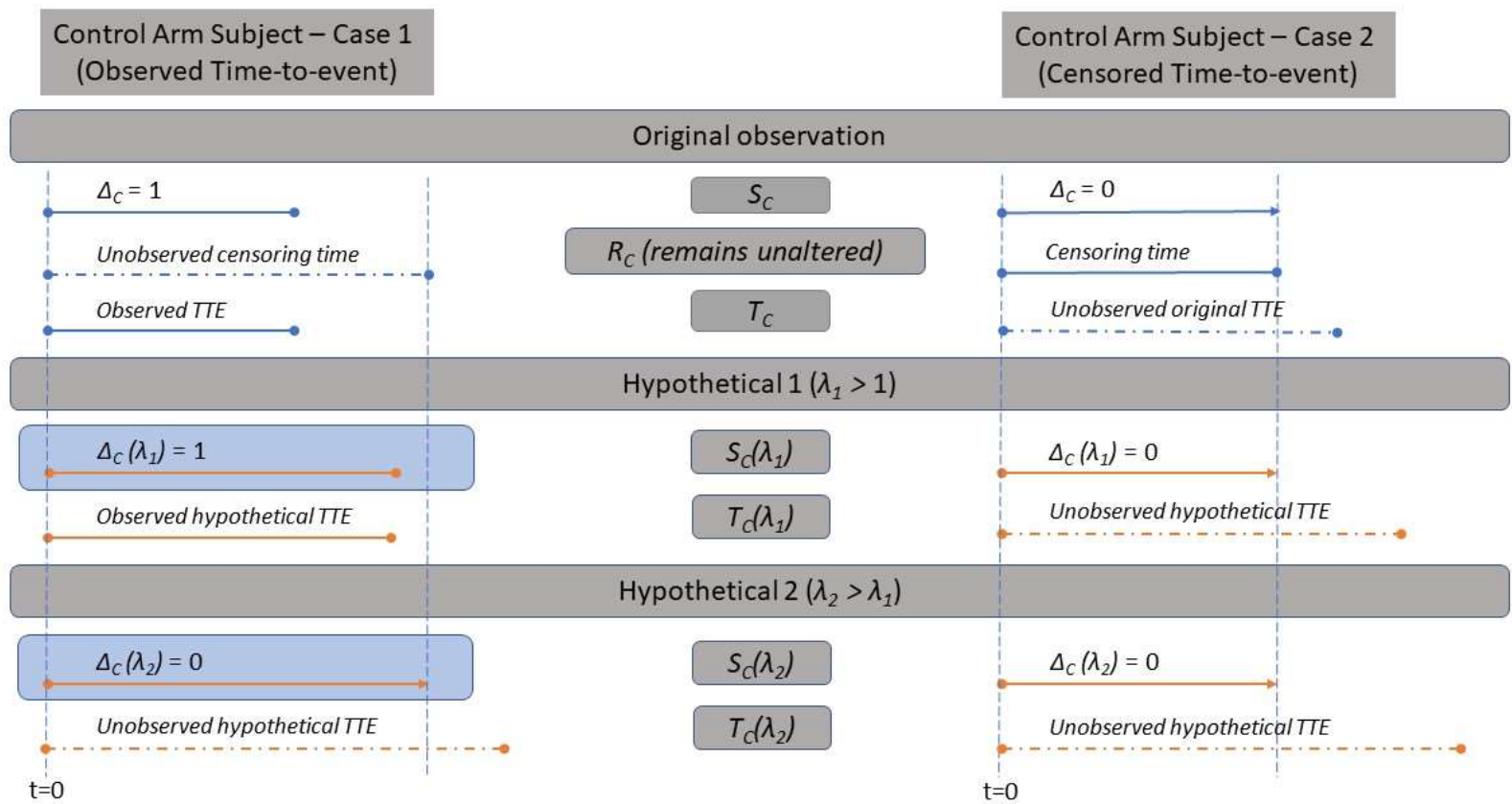

**Figure 2.** Diagram representing the results shown in Proposition 1a and counterfactual elicitation for the assessment of Effect 1.

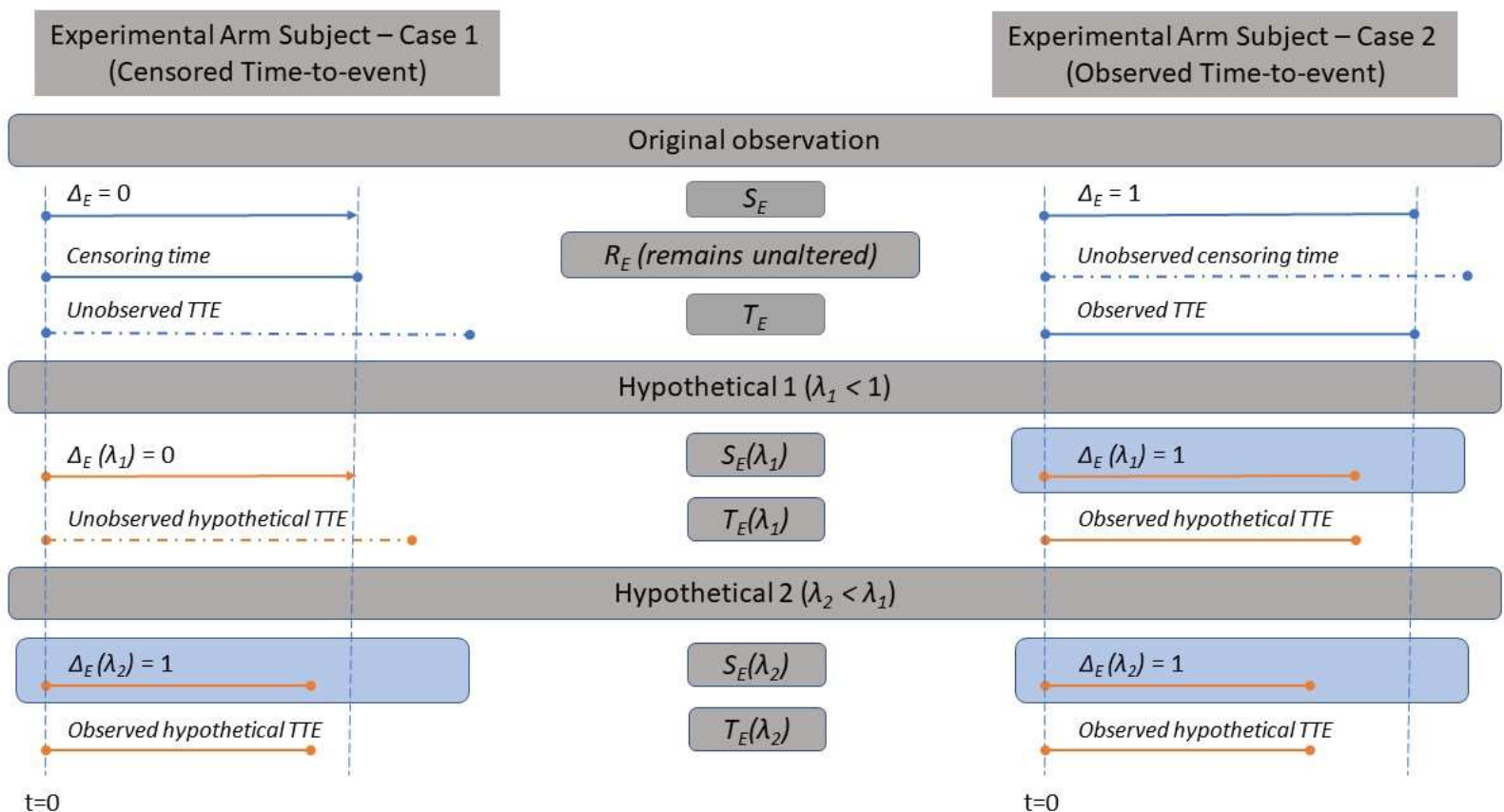

**Figure 3.** Diagram representing the results shown in Proposition 1b and counterfactual elicitation for the assessment of Effect 2.

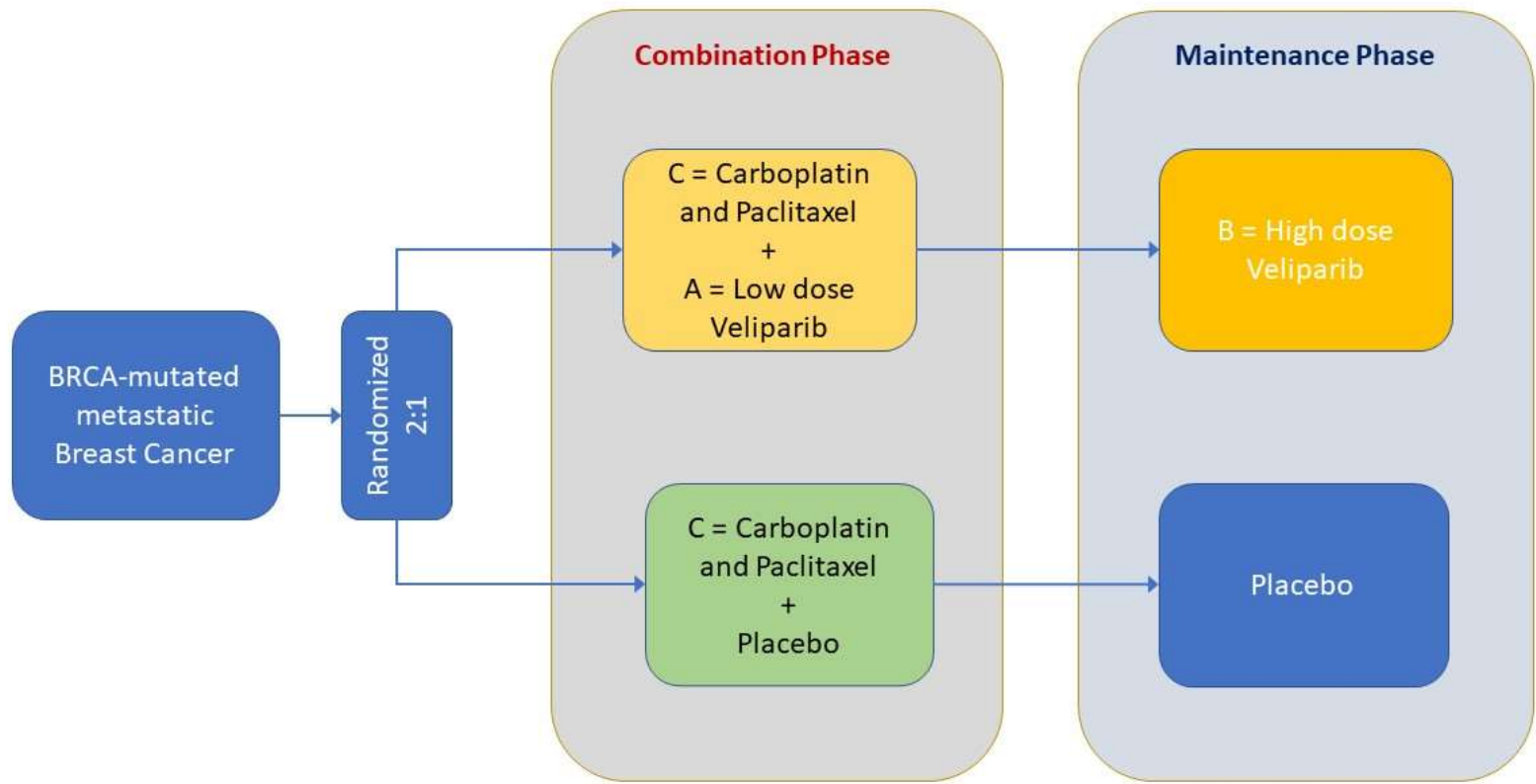

**Figure 4.**     BROCADE3 Study Design

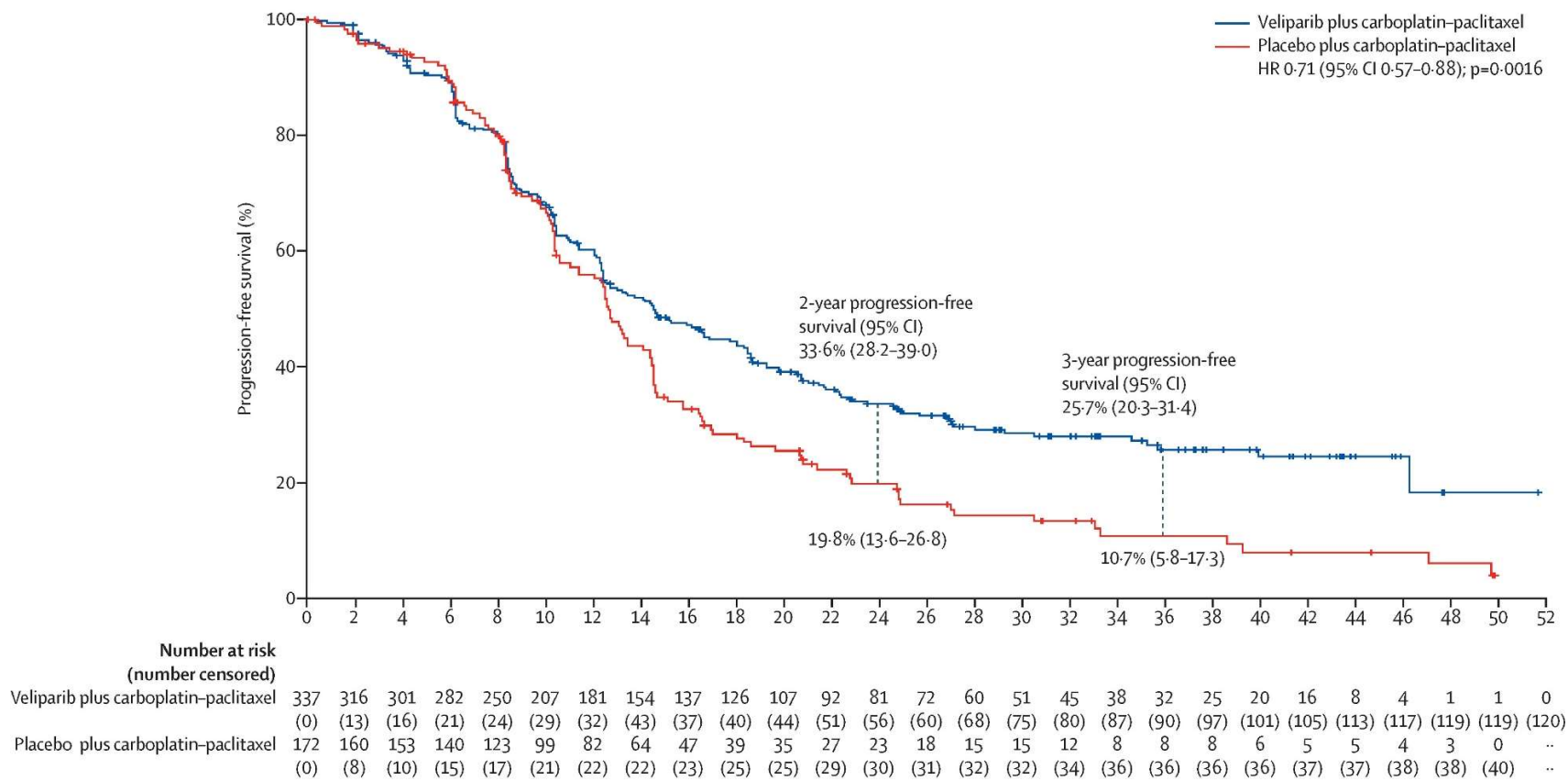

**Figure 5.** Progression-free survival results of the BROCADE3 study (reprinted with permission of Elsevier Pvt. Ltd. from article by Diéras et al. (2020)).

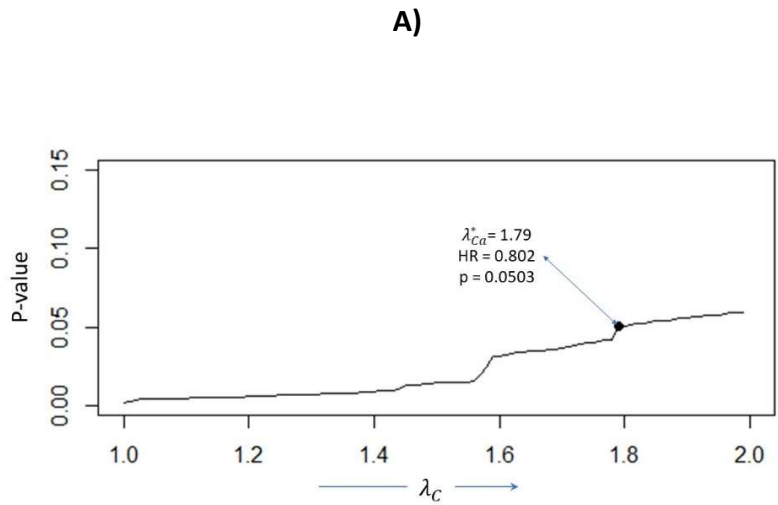
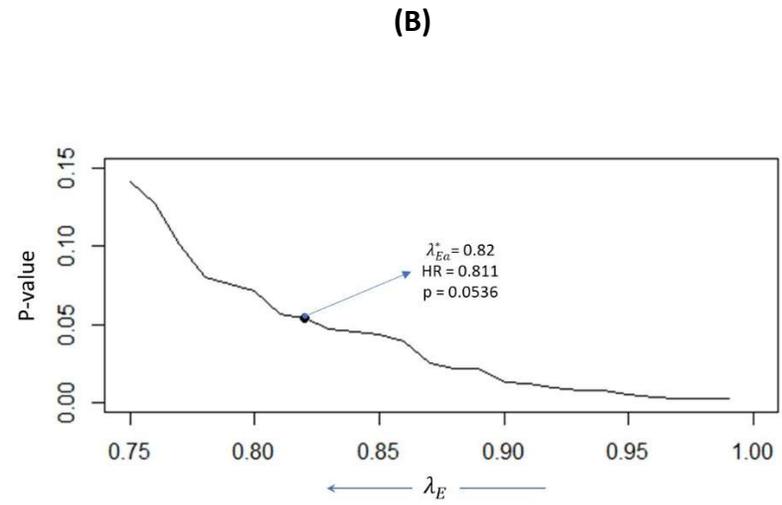
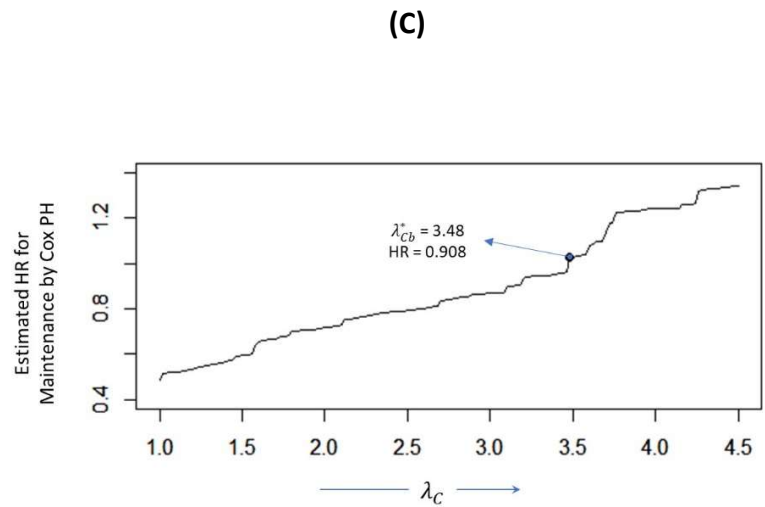
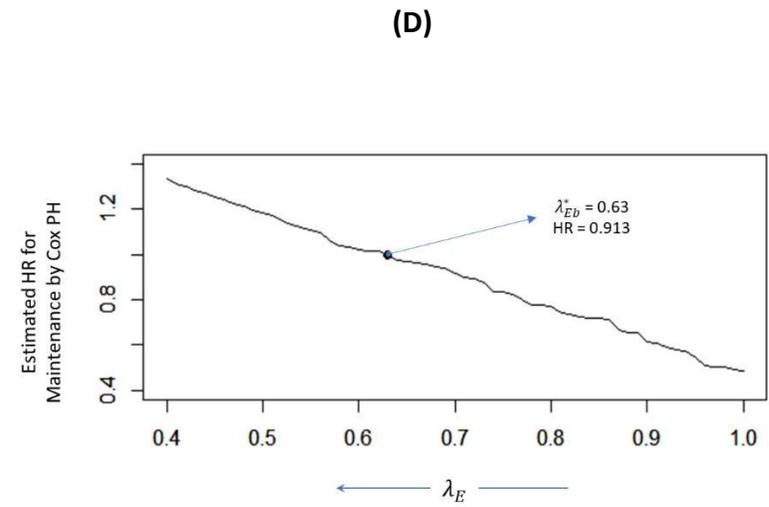

**Figure 6.** Results of tipping point analysis using RPSFT models for Effect 1: (A) and (C) and for effect 2 (B) and (D) (A) and (B): plot of counterfactual p-values vs. values of $\lambda_C$ (for effect 1) and $\lambda_E$ (for effect 2)

(C) and (D): plot of counterfactual hazard ratios during monotherapy vs. values of $\lambda$ for effect 1 and effect 2